\documentclass[conference]{IEEEtran}
\IEEEoverridecommandlockouts
\usepackage{cite}
\usepackage{comment}
\usepackage{amsmath,amssymb,amsfonts}
\usepackage{algorithmic}
\usepackage{algorithm}  
\usepackage{graphicx}
\usepackage{textcomp}
\usepackage{xcolor}
\usepackage{float} 
\usepackage{color}
\usepackage{multirow}
\usepackage{array}
\usepackage{booktabs}
\usepackage{multirow}
\usepackage{fancyhdr} 

\def\BibTeX{{\rm B\kern-.05em{\sc i\kern-.025em b}\kern-.08em
    T\kern-.1667em\lower.7ex\hbox{E}\kern-.125emX}}
    
    \newcommand{\notered}[1]{\textcolor{red}{[{\bf #1}]}}

\begin{document}

\title{ Correlated  Deep Q-learning based Microgrid Energy Management\\
}
\author{\IEEEauthorblockN{Hao Zhou and Melike Erol-Kantarci, \IEEEmembership{Senior Member, IEEE}}
\IEEEauthorblockA{\textit{School of Electrical Engineering and Computer Science} \\
\textit{University of Ottawa}\\
Emails:\{hzhou098, melike.erolkantarci\}@uottawa.ca}}
\maketitle
\thispagestyle{fancy} %
      \lhead{} 
      \chead{Accepted by 2020 IEEE 25th International Workshop on CAMAD, 978-1-7281-6339-0/20/\$31.00 ©2020 IEEE
 } 
      \rhead{} 
      \lfoot{} 
      \cfoot{\thepage} 
      \rfoot{} 
      \renewcommand{\headrulewidth}{0pt} 
      \renewcommand{\footrulewidth}{0pt} 
\pagestyle{fancy}
\cfoot{\thepage}
 
\begin{abstract}
Microgrid (MG) energy management is an important part of MG operation. Various entities are generally involved in the energy management of an MG, e.g., energy storage system (ESS), renewable energy resources (RER) and the load of users, and it is crucial to coordinate these entities. Considering the significant potential of machine learning techniques, this paper proposes a correlated deep Q-learning (CDQN) based technique for the MG energy management. Each electrical entity is modeled as an agent which has a neural network to predict its own Q-values, after which the correlated Q-equilibrium is used to coordinate the operation among agents. In this paper, the  Long Short Term Memory networks (LSTM) based deep Q-learning algorithm is introduced and the correlated equilibrium is proposed to coordinate agents. The simulation result shows 40.9\% and 9.62\% higher profit for ESS agent and photovoltaic (PV) agent, respectively.   
\end{abstract}
    

\section{Introduction}
The integration of microgrids (MGs) to the electrical distribution system will improve the efficiency, reliability and resilience of future smart grid \cite{b1}. MGs are electrical systems that can connect to the main grid or work in island-mode. In particular, they bare many opportunities for enhancing the reliability of the electricity grid when they are able to form energy sharing communities \cite{b2, b3}. 

An MG usually contains an energy storage system (ESS), one or more generators which can be in the form of renewable energy resources (RER) and the load of users. MGs can be modelled as multi-agent systems where storage, generation and load are each represented by an agent and the agents might have different owners or controllers. For example, a RER unit can sell energy to the grid or to the ESS. The decision on who to sell and the price of selling, will determine the revenue for the RER agent. Cooperation and competition may exist between agents to maximize their own utility or the overall profit\cite{b4}. In this case, the agents need to trade energy to maximize their own profit, and the competition may increase the difficulty of balancing the profit. To this end, considering the potential of machine learning techniques learning based methods are promising to help agents maximize their profit. 

The power loss during MG energy trading is considered in \cite{b5}, and a Bayesian based reinforcement learning technique is applied for the optimal energy trading strategy among MGs. Furthermore, the environment in a MG is related with the state of each agent, which means the MG system state could be rather complicated. Deep reinforcement learning (DRL) \cite{b6} is applied in the MG energy management field in several prior works.
The DRL is used for Peer-to-Peer energy trading in\cite{b7}, and an improvement is shown when compared with rule-based strategy. Furthermore, energy management problem is formulated to be a markov decision process (MDP) in \cite{b8}, and actor-critic based DRL is proposed to minimize the energy cost. In addition, considering the overestimation of standard deep Q-learning, the double deep Q-learning (DQN) is used in \cite{b9} for community ESS, and the result proves the feasibility of double DQN.

Most of the works in the literature \cite{b6,b7,b8,b9} mainly focus on one-agent case. However, the decentralized method is more suitable and scalable for MG application \cite{b10}. A multi-agent reinforcement learning (MARL) method is proposed in \cite{b11}, where each agent has the learning ability to maximize its own profit. In addition, the cooperative reinforcement learning is proposed in \cite{b12} for the MG economic dispatch, where the cooperation is guaranteed by the diffusion strategy. In MARL, each agent is expected to learn the cooperation strategy. Naturally, MARL can also be generalized to multi-agent deep reinforcement learning (MADRL). Nash based method are generally used to distribute profits among cooperative agents \cite{b13}. However, it needs to exchange information iteratively, which may lead to a heavy computation burden. In this paper, we use the correlated equilibrium (CE) to harmonize agents. In CE, the agents exchange their Q-values, and make an optimal joint action \cite{b14}. The CE is conducted by a linear program, which is easy to be implemented in a decentralized way.   

In \cite{b15}, the CE based reinforcement learning is used for dynamic transmission control of sensor network, and the results show that CE can balance agents in a distributed manner. Combined with on policy SARSA($\lambda$) algorithm, the CE is applied for the smart generation control of interconnected grid in \cite{b16}. However, the main difference between this paper and other related work is that, we propose a correlated deep Q-learning method for MG energy management, which is a MADRL algorithm. To the best of our knowledge, this is the first time CDQN is proposed and applied in MG field.

The main contribution of this paper is that we generalize MARL to multi-agent deep reinforcement learning. We propose a correlated deep Q-learning method for the MG energy management, where each agent runs the DQN independently and the CE is used for coordination. The decentralized manner aims to protect agents' privacy. Our simulation results demonstrate the success of CDQN by having 40.9\% and 9.62\% higher profit for ESS agent and PV agent, respectively. 

The reminder of this paper is organized as follows. Section II defines the agents energy model and MG energy trading model. Section III presents the problem formulation. Section IV introduces the Q-learning, deep Q-learning, and correlated equilibrium. Section V is the simulation result. Section VI concludes this paper.

\section{Microgrid agents and energy trading model}
The energy models of DSM agent,  ESS agent and PV agent are introduced in this section. Then the model for energy trading is presented.

\subsection{DSM Agent Model}
In an MG, the end user's devices can be generally divided as crucial load and deferrable load. The crucial load mainly includes light and cooking devices, which are unable to change their operation time. However, the operation time of deferrable load could be changed without affecting the end users' satisfaction much, e.g., washing machine. We assume the DSM agent is authorized by consumers to control the deferrable load of the MG. 

Let $\vec P=[P_{1},P_{2},...,P_{D}]$ be the power demand of $D$ deferrable devices, and $\vec o_{t}=[o_{t,1},o_{t,2},\dots, o_{t,D}]$ is the working state (on/off) of devices at time $t$. Then the total aggregated power demand of devices at time $t$, constitute the power demand of the DSM agent at time $t$: 
\begin{equation} \label{eu_eqn}
P^{DSM}_{t} =\sum_{p=1}^{D}P_{p}o_{t,p}
\end{equation}
where $o_{t,p}=1$ denotes the device is on, and $o_{t,p}=0$ denotes it is off.

Meanwhile, the deferrable devices should be serviced before a certain time limit, i.e. the waiting time should be bounded, such that user comfort is not jeopardized. The waiting status is the maximum allowable waiting time of $n$ deferrable devices and is denoted as $\vec w_{t}=[w_{t,1},w_{t,2},\dots, w_{t,n}]$. The devices have to be serviced before $w_{t,p}=0$, or a high penalty will be applied for affecting the end user's comfort.

\subsection{Energy Storage System Model}
In this paper, the ESS is assumed to be a single unit, even through multiple storage devices can exists. They are assumed to be controlled by the same agent. To simplify the model, we assume the ESS agent charge and discharge with a fixed power rate $P^{ch}$. The ESS energy model is described as:
\begin{equation} \label{eu_eqn}
P^{ESS}_{t} =P^{ch}z_{t}
\end{equation}
where $z_{t}=-1$ denotes charge, $z_{t}=0$ denotes unchanged, and $z_{t}=1$ denotes discharge. 

The state of charge (SOC) of ESS is updated according to: 
\begin{equation} \label{eu_eqn}
SOC_{t+1} =SOC_{t}-\frac{P^{ch}}{C^{ESS}}z_{t}
\end{equation}
where $C^{ESS}$ is the capacity of ESS.

\subsection{PV Model}
In this paper, we assume the PV power can be  predicted based on prior works in the literature \cite{b17}.
\begin{equation} \label{eu_eqn}
P^{PV}_{t} =\hat{P}^{PV}_{t}
\end{equation}

\subsection{Microgrid Energy Trading Market}
In this section, we present an auction based energy trading market \cite{b18} to set the price of electricity in each transaction. There are two kinds of participants in this market, which are suppliers and consumers. For example, in our problem, the PV agent works as a supplier, and DSM agent works as a consumer. The ESS agent could be supplier when discharging, and a consumer when charging. 
\begin{figure}[h]
\centering
\includegraphics[width=6cm,height=3.5cm]{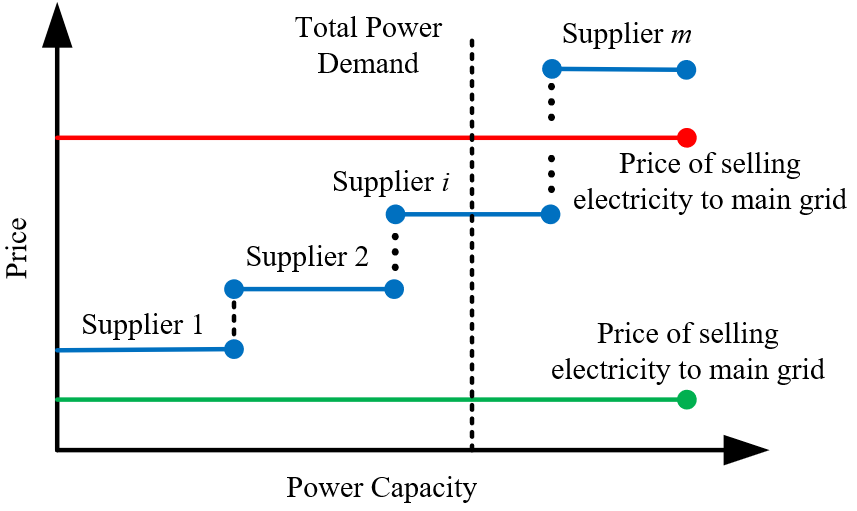}
\caption{Auction based energy trading.}
\label{fig}
\end{figure}

Let's assume there are $n$ suppliers and $m$ consumers in the MG. At each time slot $t$, the consumers will submit their own power demand, and the total power demand is calculated. For the suppliers, they need to submit their power supply and bidding price. 

As shown in Fig.1, the suppliers will be ranked from lowest price to highest price. When the total power supply is higher than total power demand, it is assumed that the supplier $i$ is the intersection of total power demand and supply. Then the bidding price of supplier $i$ becomes the uniform clearing price for supplier 1 to $i$. The rest of the power supply will be sold to the main grid, which is usually a lower price. On the other hand, when the total power demand is higher than total power supply, the main grid will participate the trading as a supplier, and the price of buying electricity from the main grid becomes the uniform clearing price for all suppliers. Under this trading market model, the suppliers will compete for a lower price to reduce the possibility of selling electricity to the main grid, which will benefit the MG consumers. Note that the bidding price should between the lower and upper bound given by main grid, or the consumers will prefer to buy electricity from main grid, or supplier will sell electricity to the main grid.  

\section{Problem Formulation }
The objectives and constraints of each agent are defined in this section. 

The DSM agent always works as a consumer and aims to minimize the cost:
\begin{equation} \label{eu_eqn}
min(\sum^{T}_{t=1}\sum_{p=1}^{D}P_{t}^{DSM}p^{c}_{t})
\end{equation}
where $T$ is the simulation horizon, $p^{c}_{t}$ is the clearing price at time $t$, which is calculated by the energy trading model. 

The PV agent aims to maximize its utility:
\begin{equation} \label{eu_eqn}
max(\sum^{T}_{t=1}({P}^{PV}_{t}p^{s}_{t}-C_{PV}(P_{t}^PV)))
\end{equation}
where ${P}^{PV}_{t}$ is PV power, and $p^{s}_{t}$ is the average selling price for PV power at time $t$. $C_{PV}$ is the cost function, which is related with the PV generation efficiency.

ESS agent could be a supplier when discharging or a consumer when charging. It aims to maximize its utility by:
\begin{equation} \label{eu_eqn}
max(\sum^{T}_{t=1}P^{ESS}_{t}p^{tra}_{t})
\end{equation}
where $p^{tra}_{t}$ is the average trading price for ESS at time $t$. $P^{ESS}_{t}$ will be a negative value when charging, which means ESS agent desires a lower price when charging, and higher price when discharging. 

The problem is optimized under following constraints:
\begin{equation} \label{eu_eqn}
P^{PV}_{t}+P^{grid}_{t}+P^{ESS}_{t}=P^{DSM}_{t}
\end{equation}
\begin{equation}
o_{t,i}\leq w_{t,i}
\end{equation}
\begin{equation} \label{eu_eqn}
SOC_{min}\leq SOC_{t} \leq SOC_{max}
\end{equation}
\begin{equation} \label{eu_eqn}
p_{min}\leq p_{t} \leq p_{max}
\end{equation}
Equation (8) is the energy balance constraint, where $P^{grid}_t$ is the power of the main grid. Equation (9) is the DSM constraint, which means only devices that have not been serviced could be turned on. Equation (10) is the SOC constraint, where $SOC_{min}$ and $ SOC_{max}$ are lower and upper bound of ESS battery. Equation (11) means the bidding price is between the lower and upper limits.  

\section{Correlated Deep Q-learning Algorithm (CDQN)}
In this section, we introduce the CDQN, where each agent uses a neural network to predict its Q-values, and correlated equilibrium is applied for the coordination of agents such that the objectives defined in the previous section is met.

\subsection{Q-learning}
Q-learning is a widely used reinforcement learning method. Based on the markov decision process (MDP), the elements of Q-learning could be described as state $s$, action $a$, reward $r$ and transition probability $T$. We will first define the $<S,A,T,R>$ for different agents.

The state and actions of DSM agent are defined as $s^{DSM}=\{t,\vec w\}$ and $a^{DSM}=\{P^{DSM}\}$, respectively, where the $\vec w$ is the waiting status.
The state of PV agent is $s^{PV}=\{t\}$, and the action is $a^{PV}=\{\hat{P}^{PV},p^{s}\}$, where $\hat{P}^{PV}$ is predicted PV power, and $p^{s}$ is the bidding price. For the ESS agent, the state is $s^{ESS}=\{t,SOC\}$, and the action is $a^{ESS}=\{P^{ESS},p^{s}\}$ when discharging and $a^{ESS}=\{P^{ESS}\}$ when charging. 

The Q-learning algorithm will choose the best action $a$ under a certain state $s$, which aims to maximize the long term total expected reward. For one agent, the Q-values are updated according to:
\begin{equation} \label{eu_eqn}
Q_{\pi}(s, a) = Q_{\pi}(s, a)+\alpha( r(s, a)+ \gamma V_{\pi}(s')-Q_{\pi}(s, a))
\end{equation}
To balance the exploitation and exploration, the $\epsilon$-greedy policy is generally applied: 
\begin{equation} \label{eu_eqn}
\pi_{i}(s) =\left\{
\begin{array}{rcl}
  randomly &   rand\leq\epsilon\\
\arg max(Q(s,a))  &  rand>\epsilon\\
\end{array} \right.
\end{equation}
where $\epsilon$ is a small number, $rand$ is a random value. When $rand\leq\epsilon$, the agent tends choose actions randomly for exploration. On the other hand, it will choose the action of max Q-value for exploitation.   

\subsection{Deep Q-learning}
In tabular Q-learning, a Q-table is used to record the accumulated reward of each action $a$ under every state $s$, and new action is chosen according to their Q-value. However, if the state-action space is large, there will be a large Q-table, which means the memory and time complexity will be high, and the algorithm will take a long time to converge. 

To this end, deep Q-learning is proposed, where the Q-values are generated by a neural network as in \cite{b6}. As shown in equation (12), when the Q-values converge, the current Q-value $Q_{\pi}(s, a)$ is equal to $r(s, a)+ \gamma V_{\pi}(s')$. Accordingly, the $r(s, a)+ \gamma V_{\pi}(s')$ is regarded as the training target of neural network, and $Q_{\pi}(s, a)$ is the prediction result. Then, we have the loss function as:  
\begin{equation} \label{eu_eqn}
L(w) = E(r(s,a)+ \gamma V(s',w)-Q(s,a,w))
\end{equation}
where $w$ is the weight of neural network, $s$ and $s'$ is the current and next state, respectively.

Q-values change dynamically in the Q-learning, and the target values also changes, which may lead to an unstable result. Meanwhile, the training data is expected to be discontinuous, but the state and action transition is consecutive in Q-learning. To improve the DQN, two improvements are proposed, namely experience replay and target network \cite{b6}.



In this paper, we use the LSTM network to predict the Q-values, which is a special recurrent neural network (RNN)\cite{b19}. LSTM network could better capture the long-term dependencies than traditional RNN network, and the vanishing gradient problem is avoided. This approach has been employed to resource allocation problems in wireless networks \cite{b20}.
\begin{figure}[!h]
\centering
\includegraphics[width=6cm,height=3.5cm]{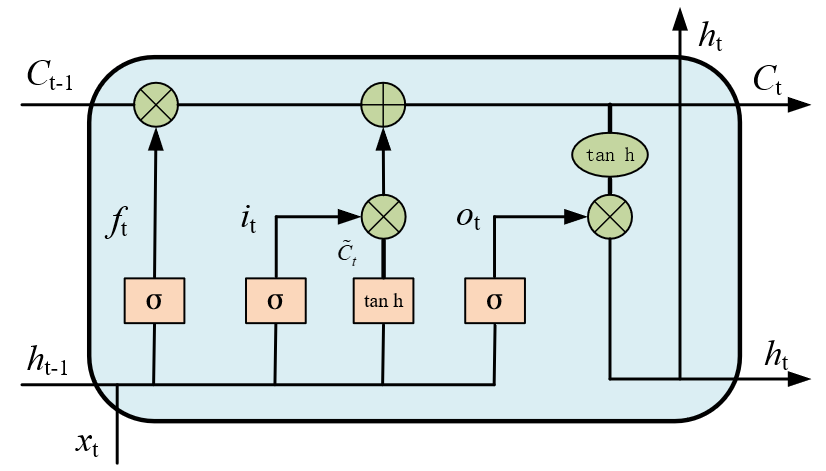}
\caption{LSTM network structure}
\label{fig}
\end{figure}
As shown in Fig.2, $x_{t}$ is the input value at time $t$, $h_{t-1}$ is the output value at time $t-1$, and $c_{t-1}$ is the cell state at time $t-1$. $h_{t}$ is the current output, and $c_{t}$ is the current cell state. 

\subsection{Correlated Equilibrium}
In one agent case, the DQN can still use the $\epsilon$-greedy approach to choose actions. However, in a multi-agent case, the coordination is necessary. In this paper, the correlated equilibrium is implemented to coordinate the action of agents. In correlated equilibrium, the coordination of agents is achieved by exchanging Q-values. After exchanging Q-values, each agent will choose actions according to the following equations:
\begin{equation}
\begin{aligned}
\max \sum_{\vec a \in A}Pr(s,\vec a, w)  Q(s,\vec a, w) \quad \quad \quad   \quad \quad\\
sub. to \sum_{\vec a \in A}Pr(s,\vec a, w)=1 \quad \quad \quad \quad \quad \quad \quad\\
\sum_{\vec a_{-i} \in A_{-i}}Pr(s,\vec a_{i}, w)(Q(s,\vec a_{i}, w)-Q(s,\vec a_{-i},a_{i}, w))\geq0\\
0 \leq Pr(s,\vec a, w)\leq 1 \quad \quad \quad \quad \quad \quad \quad\\
\end{aligned}
\end{equation}
where $\vec a$ is the joint action of agents, $Pr(s,\vec a, w)$ is a vector of the probability of $Q(s,\vec a, w)$, $\vec a_{-i}$ is the joint action except agent $i$, $a_{i}$ is the action of agent $i$, $A_{-i}$ is the set of joint action except agent $i$. Equation (15) aims to maximize the total expected reward of multi-agents, and find the probability distribution of action combination to guarantee each agent makes a optimal action. 
\begin{figure}[!h]
\centering
\includegraphics[width=7cm,height=4cm]{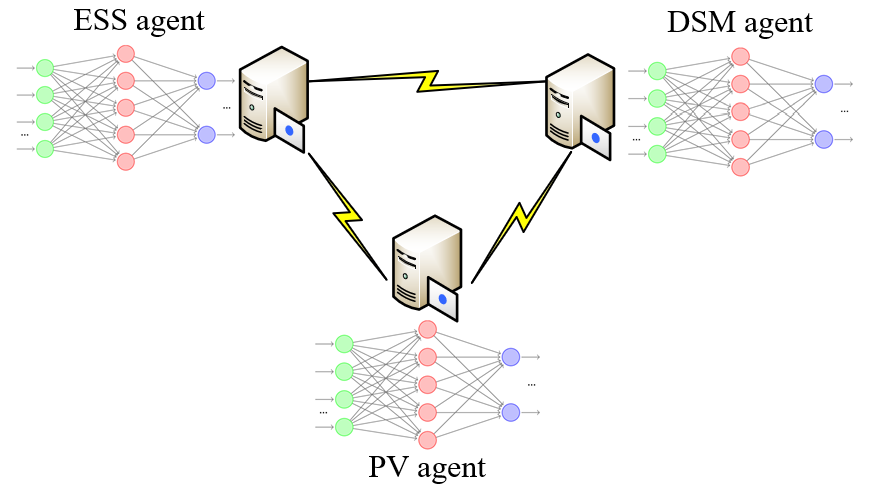}
\caption{System structure of correlated deep Q-learning.}
\label{Fig.3}
\end{figure}
As shown in Fig.3, it is worth noting the correlated equilibrium is achieved in a decentralized manner, and no central controller is needed. The state, action and reward of each agent are regarded as private information. The deep Q-learning works independently for each agent to predict the Q-values, which means each agent has its own neural network and experience pool. The only information they need to exchange is the Q-values. This decentralized scheme will help protecting the agents' privacy. 

The proposed CDQN algorithm is shown in Algorithm 1. Briefly, based on the $\epsilon$-greedy policy, when the random number is smaller than $\epsilon$, agents choose actions randomly. Otherwise, agents will exchange Q-values and make a joint action based on correlated equilibrium. After the action $\vec a$ is conducted, each agent will save $(s_{i},\vec a, r,s_{i}')$ to its own experience pool, which will be used for its own DQN training.

\begin{algorithm}[!h]
	\caption{CDQN}
	\begin{algorithmic}[1]
		\STATE \textbf{Initialize:} MG and DQN parameters
		\FOR{$j=1$ to $episode$}
		\STATE Reset state $s$ 
		\FOR{$t=1$ to $T$}
		\IF{$rand<\epsilon$}
		\STATE Each agent chooses actions randomly.  
		\ELSE
		\STATE Each agent predicts its $Q(s_{i},\vec a,w)$ for possible joint action $\vec a$ under current state $s_{i}$.
		\STATE Agents exchange Q-values under their own state $s_{i}$. Each agent finds equilibrium by equation (15) and get optimal joint action $\vec a$.
		\ENDIF
		\STATE $\vec a$ is conducted, and each agent calculates its reward $r_{i}$, and updates its own state $s_{i}$.
		\STATE Each agent saves $(s_{i},\vec a, r,s_{i}')$ to their own experience pool.
		\ENDFOR
		\STATE  Every $C$ episodes, random sample minibatch from experience pool, each agent trains its own $w$.
		\STATE Every $mC$ episodes, copy $w$ to $w'$.
		\ENDFOR
	\STATE \textbf{Output:}Optimal action sequence from $t=1$ to $T$	
	\end{algorithmic}
\end{algorithm}

\section{Simulation Results}
\subsection{Parameter Settings}
In an MG, there are PV agents, DSM agents and ESS agents. They will exchange energy in the energy trading market. The predicted PV power is shown in Fig.4. The cost of PV agent is 1.14 \$/h when PV power output is not zero. The price of buying and selling electricity with main grid is shown in Fig.5, where the price of selling electricity to main grid is set to be 50\% of buying electricity from main grid\cite{b21}. The bidding price is discretized as [0.06, 0.09, 0.12, 0.15, 0.18, 0.21] \$/kW·h.

\begin{figure}[!h]
\centering
\includegraphics[width=6.5cm,height=3.5cm]{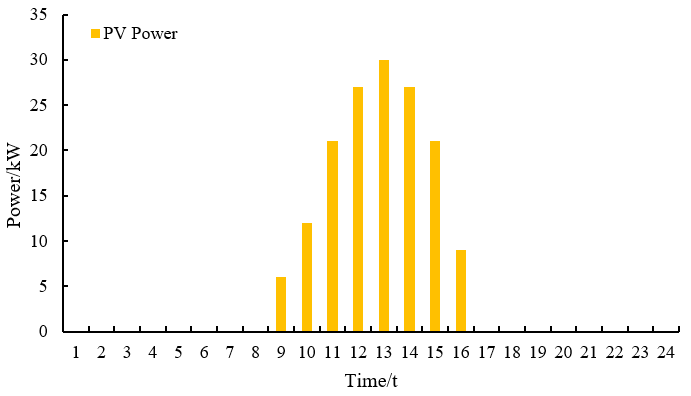}
\caption{PV output power.}
\label{fig}
\end{figure}

\begin{figure}[!h]
\centering
\includegraphics[width=6.5cm,height=3.4cm]{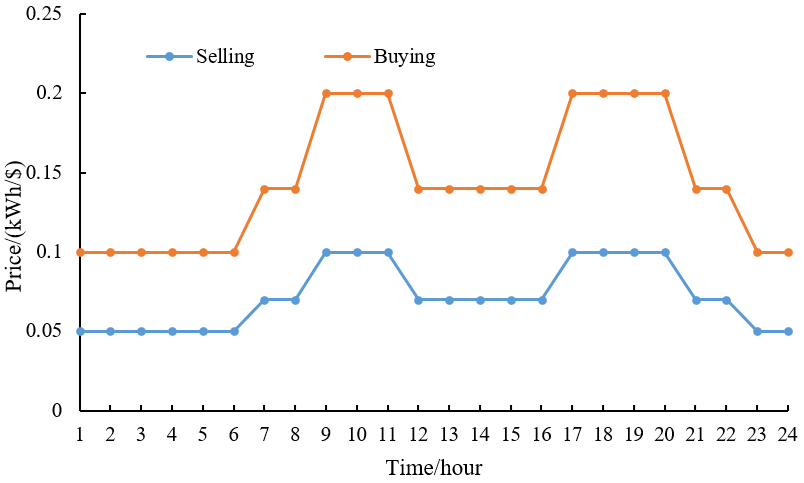}
\caption{Time of use (TOU) electricity price.}
\label{fig}
\end{figure}

\begin{table}[!hbtp]
\caption{Properties of deferrable devices}
\centering
\renewcommand\arraystretch{1.3}
\begin{tabular}{|p{1cm}<{\centering}|p{1cm}<{\centering}|p{2cm}<{\centering}|p{2cm}<{\centering}|}
\hline
\shortstack{Device\\ Number} &\shortstack{Average\\ Power}  &\shortstack{Operation time \\limit (Hours)} & \shortstack{Average duration \\time (Hours) }\\
\hline
1 &17 &[1, 8]& 1\\
\hline
2 &15 &[7, 13]& 1\\
\hline
3& 15 &[10, 17]& 1\\
\hline
4& 16 &[15, 22]& 1\\
\hline
5& 20 &[20, 4$^{\mathrm{+24h}}$]& 1\\
\hline
\end{tabular}
\end{table}

\begin{table}[!hbtp]
\caption{DQN parameters}
\centering
\renewcommand\arraystretch{1.3}
\begin{tabular}{|p{4cm}<{\centering}|p{2.5cm}<{\centering}|}
\hline
\shortstack{Parameters}   &\shortstack{Value} \\
\hline
Hidden layers  & 2 Lstm Layers \\
\hline
Hidden layers nodes  & 30 each \\
\hline
Learning rate & 0.000 6\\
\hline
Discount rate & 0.95\\
\hline
Capacity of experience pool & 1200\\
\hline
Training batch size&120\\
\hline
Training Frequency & Every 40 episodes
\\
\hline
\end{tabular}
\end{table}

The capacity of ESS is 100 kW·h, and charging power is 20kW, and initial SOC of ESS is set to 0.2. There are 5 sets of deferrable devices, where the operation time limit and average duration time is shown in Table.1. Here, $4^{+24h}$ means, 4 am of the next day. The parameters for DQN is also shown in Table 2, which is selected according to our simulation simulation performance. 

\subsection{Performance Evaluation}
The proposed CDQN algorithm is evaluated with respect to convergence and profit/cost performance. In addition, it is compared to a DQN algorithm where there is no coordination among agents. As shown in Fig.6, the profit of PV and ESS agent increase with iterations, and converge at last. A decreasing DSM cost is also observed.
\begin{figure}[!h]
\centering
\includegraphics[width=6cm,height=4.5cm]{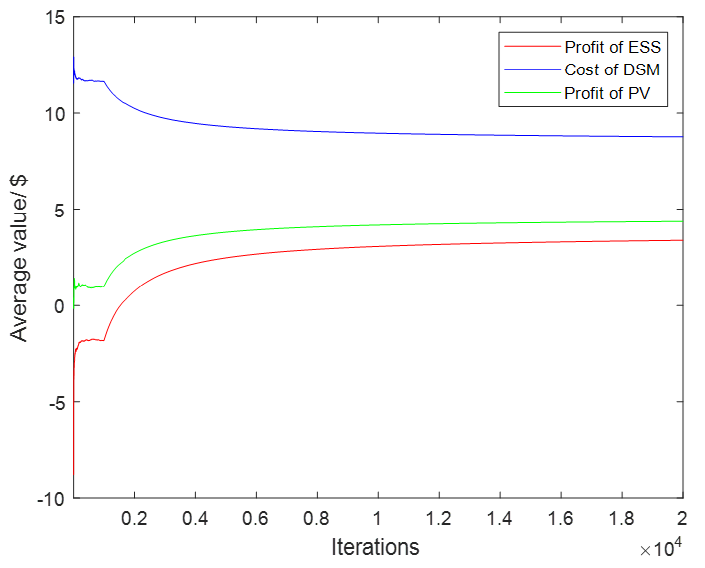}
\caption{Average profit/cost of agents.}
\label{fig}
\end{figure}

The optimal energy management strategy obtained by CDQN is shown in Fig.7. This figure gives an insight on energy trading decisions of the agents. The ESS agent uses PV power to charge at time 12, 13 and 14, and the stored electricity is sold to DSM devices at time 20 and 21. Due to a higher buy back price offered by main grid. ESS agent also sells some energy to main grid at time 18 and 19. Meanwhile, the DSM agent uses PV power for operation at time 11 and 15. DSM agent buys electricity from ESS agent at time 21 and 22, which is a lower price compared with the price offered by main grid. It's worth noting the ESS agent did not sell electricity to DSM agent at time 1, and the main reason is the trading price is too low. 
\begin{figure}[!h]
\centering
\includegraphics[width=7cm,height=4cm]{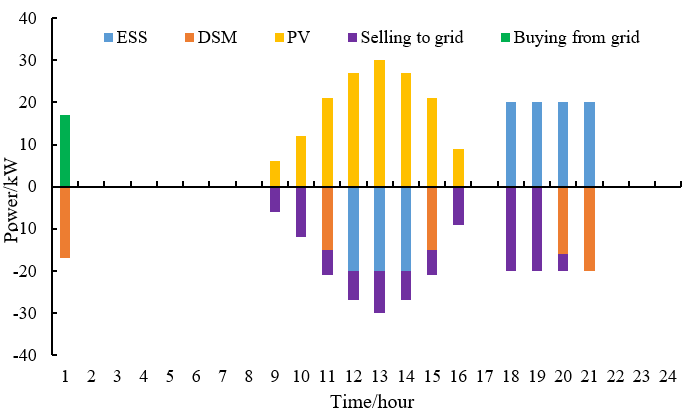}
\caption{Optimal energy trading under CDQN.}
\label{fig}
\end{figure}

The Fig.8 presents the uniform clearing price for the market, where the trading price is always between buying and selling price offered by main grid. Note that the clearing price overlaps with grid selling price at time 9, 10, 16, 18 and 19, which means all the electricity is sold to main grid. 
\begin{figure}[!h]
\centering
\includegraphics[width=7cm,height=4cm]{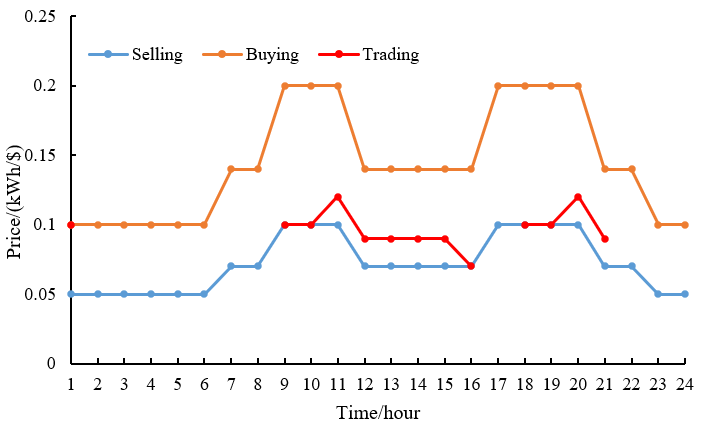}
\caption{Trading price in the MG energy market.}
\label{fig}
\end{figure}

The operation time of deferrable devices is presented in Fig.9. Devices in categories 2 and 3 are deferred 4 and 5 hours, respectively, to use the lower price PV power. Device categories 4 and 5 are deferred 5 and 1 hours, respectively, to buy electricity from the ESS agent. Device 1 is not deferred because the operation limit is in the lowest price period.     
\begin{figure}[!h]
\centering
\includegraphics[width=8.5cm,height=2.3cm]{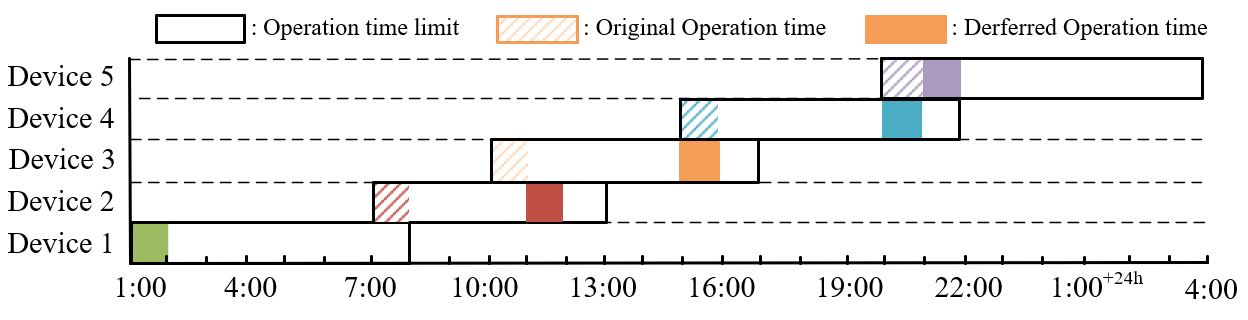}
\caption{Device operation times.}
\label{fig}
\end{figure}

We compare the performance of the proposed CDQN algorithm with a DQN algorithm where each agent works independently and only maximizes its own reward. As shown in Fig.10, a lower PV and ESS profit, and a comparable DSM cost are observed. The main reason for lower benefit for DQN is that each agent only maximize its own profit and it lacks coordination. The converged result shows ESS and PV agent have 40.9\% and 9.62\% higher profit with the proposed CDQN technique. Note that the negative values at the first 1000 iterations are due to the exploration period of the learning algorithm. 

\begin{figure}[htbp]
\centering
\includegraphics[width=6cm,height=4.5cm]{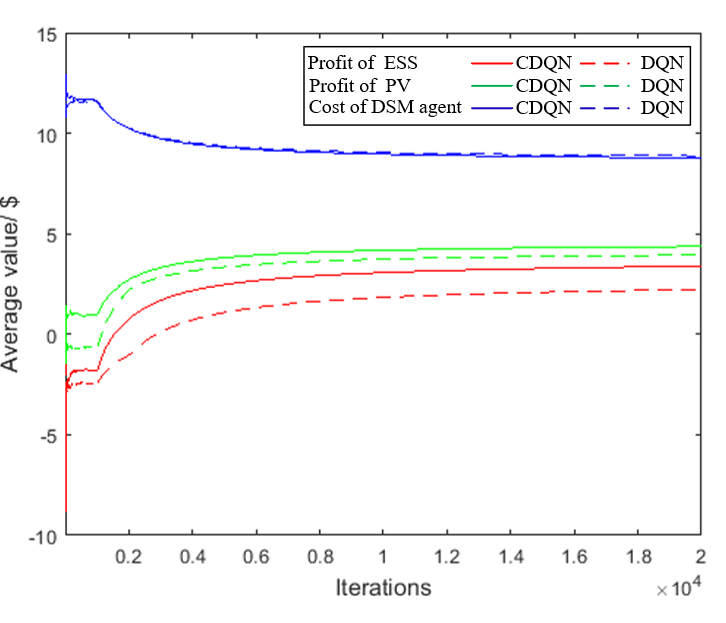}
\caption{Comparison of CDQN with DQN.}
\label{fig}
\end{figure}

\section{Conclusion}
The increasing complexity of energy trading within a microgrid and among several microgrids, call for novel approaches that benefit from the recent advances in machine learning. In this paper, we propose a correlated deep Q-learning (CDQN) technique for the MG energy management. In CDQN, each agent uses an LSTM network to predict their own Q-values and make decisions autonomously. Then, the correlated equilibrium is applied for coordination. Compared with the deep Q-learning without coordination, the CDQN performs 40.9\% and 9.62\% higher profit for ESS agent and PV agent, respectively.

\section*{Acknowledgment}
This work is supported by Natural Sciences and Engineering Research Council of Canada (NSERC), Collaborative Research and Training Experience Program (CREATE) under Grant 497981 and Canada Research Chairs Program.


\begin{thebibliography}{00}
\bibitem{b1}H. Farhangi, “The path of the smart grid,” IEEE Power and Energy Magazine, vol. 8, no.1, pp. 18-28, Dec. 2009. 
\bibitem{b2} M. Erol-Kantarci, B. Kantarci, and H. T. Mouftah, “Reliable overlay topology design for the smart microgrid network,” IEEE Network, vol. 25, pp. 38–43, Sep 2011.
\bibitem{b3}  L. Wu, J. Li, M. Erol-Kantarci, and B. Kantarci, “An integrated reconfigurable control and self-organizing communication framework for community resilience microgrids,” The Electricity Journal, vol. 30, pp.27-34, May 2017.

\bibitem{b4} J. Vardakas, I. Zenginis, N. Zorba, C. Echave, M. Morató, and C. Verikoukis, “Electrical Energy Savings through Efficient Cooperation of Urban Buildings: The Smart
Community Case of Superblocks’ in Barcelona,” in IEEE Communications Magazine, vol. 56, no. 11, pp. 102-109, Nov. 2018. 

\bibitem{b5}  M. Sadeghi, and M. Erol-Kantarci, “Power Loss Minimization in Microgrids Using Bayesian Reinforcement Learning with Coalition Formation,” in Proc. of IEEE Annual International Symposium on Personal, Indoor and Mobile Radio Communications, pp. 1-6, Istanbul, Turkey, Sep 2019.

\bibitem{b6} V. Mnih, K. Kavukcuoglu, and D. Silver, “Human-level control through deep reinforcement learning,” Nature, vol. 518, no. 7540, pp. 529–533, Jan. 2015.

\bibitem{b7} T. Chen, and S. Bu, “Realistic Peer-to-Peer Energy Trading Model for Microgrids using Deep Reinforcement Learning,” in Proc. of IEEE PES Innovative Smart Grid Technologies Europe, pp.1-5, Bucharest, Romania, Sep. 2019. 
\bibitem{b8} L. Yu, W. Xie, D. Xie, Y. Zou, D. Zhang, Z. Sun, L. Zhang, Y. Zhang, and T. Jiang, “Deep Reinforcement Learning for Smart Home Energy Management,”  vol.7, no.4, pp. 2751-2762, Dec. 2019. 

\bibitem{b9}V. Bui, A. Hussain, and H. Kim, “Double Deep Q-Learning-Based Distributed Operation of Battery Energy Storage System Considering Uncertainties,” IEEE Trans. Smart Grid, vol.11, no.1, pp. 457 – 469, Jan. 2020 

\bibitem{b10}L. Meng et al., “Microgrid supervisory controllers and energy management systems: A literature review,” Renew Sust. Energ. Rev., vol.60, pp. 1263-1273, Jul. 2016.

\bibitem{b11} E. Foruzan, L. Soh, and S. Asgarpoor, “Reinforcement Learning Approach for Optimal Distributed Energy Management in a Microgrid,” IEEE Trans Power Syst., vol. 33, pp. 5749-5758, Sep 2018.

\bibitem{b12}W. Liu, P. Zhuang, H. Liang, J. Peng, and Z. Huang, “Distributed Economic Dispatch in Microgrids Based on Cooperative Reinforcement Learning,” IEEE Trans. on Neural Networks and Learning Systems, vol.29, no.6, pp. 2192 – 2203, Mar. 2018.

\bibitem{b13} I. Zenginis, J. S. Vardakas, J. Abadal, C. Echave, M. Morat´o, and C. Verikoukis, “Optimal power equipment sizing and management for cooperative buildings in microgrids,” in IEEE Trans. on Industrial Informatics, vol. 15, no. 1, pp. 158-172, Jan. 2019.

\bibitem{b14}A. Greenwald, K. Hall, and M. Zinkevich, “Correlated Q-learning,” Dept. of Comput. Sci., Brown University, USA, Tech. Rep.,CS-05-08, 2005.

\bibitem{b15}J. W. Huang, Q. Zhu, V. Krishnamurthy, and T. Basar, “Distributed Correlated Q-Learning for Dynamic Transmission Control of Sensor Networks,” in in Proc. IEEE Int. Conf. on Acoustics, Speech and Signal Processing, Dallas, TX, USA, Mar. 2010.

\bibitem{b16}T. Yu, H. Z. Wang, B. Zhou, K. W. Chan, and J. Tang, “Multi- Agent Correlated Equilibrium Q($\lambda$) Learning for Coordinated Smart Generation Control of Interconnected Power Grids,” IEEE Trans. Power Syst., vol. 30, no. 4, pp. 1669–1679, Jul. 2015.


\bibitem{b17}European Network of Transmission System Operators for Electricity (ENTSO-E), [Online]. Available: https:// transparency.entsoe.eu/dashboard/show

\bibitem{b18}P. Shamsi, H. Xie, A. Longe, and J. Joo, “Economic Dispatch for an Agent-Based Community Microgrid,” IEEE Trans. Smart grid, vol.7, no.5, pp. 2317 – 2324, Oct. 2015. 

\bibitem{b19} S. Hochreiter and J. Schmidhuber, “Long short-term memory,” Neural Computation, vol. 9, no. 8, pp. 1735–1780, 1997

\bibitem{b20} M. Elsayed and M. Erol-Kantarci, “AI-Enabled Future Wireless Networks: Challenges, Opportunities, and Open Issues,” IEEE Vehicular Technology Magazine, vol. 14, no.3, pp. 70-77, Sep. 2019.  

\bibitem{b21} C. Ju, P. Wang, L. Goel and Y. Xu, “A Two-Layer Energy Management System for Microgrids With Hybrid Energy Storage Considering Degradation Costs,”  IEEE Trans. Smart Grid. Vol.9, No.6, Nov. 2018.

\end{thebibliography}
\end{document}